\documentstyle[12pt,my,epsfig]{article}   
\newlength{\dinwidth}                       
\newlength{\dinmargin}                      
\begin{document}
\newcommand{\Pom}{I$\!$P}                
\newcommand{\PYTHIA}{Jetsetc}
\newcommand{\SMALLX}{SMALLXa,SMALLXb}
\newcommand{\CCFM}{CCFMa,CCFMb,CCFMc,CCFMd}
\newcommand{\BFKL}{BFKLa,BFKLb,BFKLc}
\newcommand{\LDC}{LDCa,LDCb,LDCc,LDCd}
\newcommand{\DGLAP}{DGLAPa,DGLAPb,DGLAPc,DGLAPd}
\newcommand{\SHA}{GLR,LRSSa,off_shell_me,CE}
\newcommand{\JETSET}{Jetseta,Jetsetb,Jetsetc}
\def\CASCADE{{\sc Cascade}}
\def\SMMOD{{\sc Smmod}}
\def\SMALLX{{\sc Smallx}}
\def\lsim{\mathrel{\rlap{\lower4pt\hbox{\hskip1pt$\sim$}}
    \raise1pt\hbox{$<$}}}                
\def\gsim{\mathrel{\rlap{\lower4pt\hbox{\hskip1pt$\sim$}}
    \raise1pt\hbox{$>$}}}                
\setcounter{footnote}{0}
 LUNFD6/(NFFL-7176) 1999  \\
\vspace*{10mm}
\begin{center}  \begin{Large} \begin{bf}
Charm production in the semi-hard approach of QCD 
and the unintegrated gluon distribution  \\
  \end{bf} \end{Large}  
  \vspace*{5mm}
  \begin{large}
  S.P.~Baranov$^a$, H.~Jung$^b$, N.P.~Zotov $^c$ \\
  \end{large}
$^a$ 
{ P.N.Lebedev~Physical~Institute,~Russian~Academy~of~Science,~117924~Moscow,~Russia}\\
$^b$ 
{  Department of Physics, Lund University , 221 00 Lund, Sweden}\\   
$^c$
{ D.V.Skobeltsyn~Institute~of~Nuclear~Physics,
   M.V.Lomonosov~Moscow~State~University,~119899~Moscow,~Russia}\\
  \end{center}
\begin{quotation}
\noindent
{\bf Abstract:}
In the framework of semi-hard QCD approach, we present a consistent analysis 
of  $D^*$ meson production  at HERA
energies.  The consideration  is based on universal unintegrated
 gluon densities, which
have BFKL behavior in the small x region. Predictions  of  the CCFM 
 evolution equation for
 $D^*$ production are obtained and show a
 good description of $D^*$ data at HERA.
\end{quotation}

\section{Introduction}
 The standard parton model
 is based on the DGLAP~\cite{\DGLAP} 
evolution equations, which re-sums contributions from 
$[\alpha_s  \mbox{ln}(\mu^2/\Lambda^2)]$. It
 represents an one-dimensional phase space approximation for the
parton motion, also known as the collinear approximation, 
which gives the correct behavior of the structure functions at not too small
values of $x$.
When $x$ becomes smaller, also contributions from 
$[\alpha_s  \mbox{ln}(\mu^2/\Lambda^2)\,\mbox{ln}(1/x)]$ and
$[\alpha_s  \mbox{ln}(1/x)]$ need to be considered.
 In the  so called $k_t$ factorization 
or semi-hard approach (SHA)~\cite{\SHA}, 
the transverse momenta of the partons in the evolution from large $x$ at the
proton vertex towards small $x$ at the hard interaction vertex are taken into
account. This evolution in $x$ has been 
 formulated in terms of the BFKL~\cite{\BFKL}
 evolution equation. The 
 CCFM~\cite{\CCFM} evolution equation includes coherence
 effects via angular ordering and it reproduces the BFKL 
 (DGLAP) evolution equation in the small (large) $x$ limits, respectively.
\par
The resummation \cite{\SHA} of the terms
$[\alpha_s  \mbox{ln}(\mu^2/\Lambda^2)]$,
$[\alpha_s  \mbox{ln}(\mu^2/\Lambda^2)\,\mbox{ln}(1/x)]$ and
$[\alpha_s  \mbox{ln}(1/x)]$ in SHA results in the so called 
unintegrated gluon distribution
${\cal F}(x,k_t^2,Q^2_0)$, which determines the probability 
to find a gluon carrying the longitudinal momentum fraction $x$
and transverse momentum $k_t$. The factorization  scale $Q^2_0$ (such that
$\alpha_s(Q^2_0) < 1$) indicates the non perturbative input distribution.
They obey the BFKL~\cite{\BFKL} or 
CCFM~\cite{\CCFM} 
equation and reduce to the conventional parton densities $F(x,\mu^2)$ 
once the $k_t$ dependence is integrated out:
\begin{equation} \label{kt}
\int_0^{\mu^2}\!\!{\cal F}(x,k_t^2,Q^2_0)\;dk_t^2=x\,F(x,\mu^2,Q^2_0).
\end{equation}
However, in CCFM the unintegrated parton distribution 
${\cal A}(x,k_t^2,Q^2_0,\bar{q}^2)$ (instead of
${\cal F}(x,k_t^2,Q^2_0)$)  depends also on 
the maximum
angle for any emission corresponding to $\bar{q}$ 
(coming from angular ordering).
 In the small $x$ limit 
they reduce to 
${\cal F}$~\cite{CCFMd}.
\par
To calculate the cross section of a physical process, the unintegrated 
functions ${\cal F}$ or ${\cal A}$ have to be convoluted
with off-mass shell matrix elements \cite{off_shell_me,CE}
corresponding to the relevant partonic subprocesses.
In off-mass shell matrix element the virtual gluon polarization 
tensor is taken in form of SHA prescription \cite{GLR}:
\begin{equation}
L^{(g)}_{\mu\nu}=\overline{\epsilon_2^{\mu}\epsilon_2^{*\nu}}
  =p^\mu p^\nu x^2/|k_t|^2 
  =k_t^\mu k_t^\nu/|k_t|^2.
\end{equation}
\par
The specific properties of semi-hard theory may manifest in several ways.
With respect to inclusive production properties, one obtains an additional
contribution to the cross sections due to the integration over the $k_t^2$
region above $\mu^2$ and the broadening of the $p_t$ spectra due to extra 
transverse momentum of the interacting 
gluons~\cite{GLR,LRSSa,saleev_zotov_a,saleev_zotov_b}.
It is important that the gluons are not on-mass shell but are characterized
by virtual masses proportional to their transverse momentum. 
This also assumes a modification of the polarization
density matrix. A striking consequence of this fact on the $J/\psi$ spin
alignment has been demonstrated in \cite{baranov}.
\par
In this paper we present predictions
for the production of $D^*$ mesons in photo-production at HERA using the SHA
approach. We  use 
an  unintegrated gluon density coming from a solution of the CCFM equation (see
\cite{CASCADE}). We also show predictions based on a 
parton level Monte Carlo integration using a BFKL-like 
parameterization of the unintegrated gluon density,
 and we compare both predictions.
\section{Unintegrated gluon distribution and CCFM  evolution}

\begin{figure}[htb]
  \vspace*{2mm}
\epsfig{figure=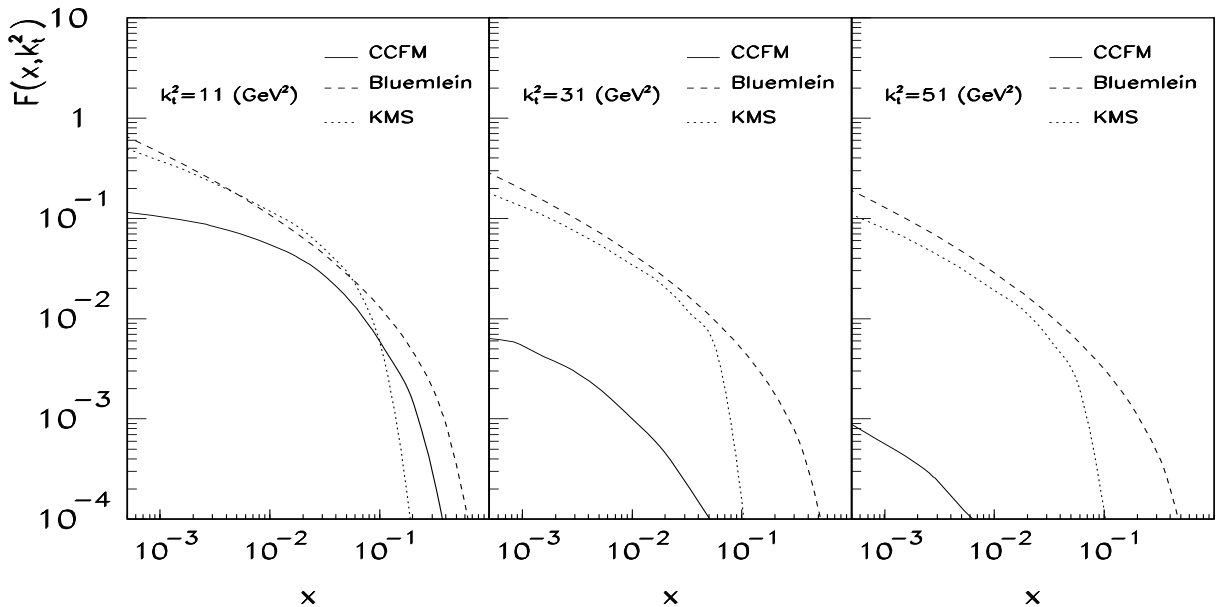,
width=17cm,height=10cm}
\caption{{\it
The gluon density ${\cal A}(x,k_t^2,\bar{q}^2)$ (solid line) obtained from
a solution of the CCFM equation~\protect\cite{CASCADE} as a function of
$x$ for different values of $k_t^2$ (at $\bar{q}^2=10$ GeV$^2$).
Also shown for comparison is the unintegrated gluon density function
${\cal F}(x,k^2_{t},\mu^2)$
 according
to the parameterization of Bl\"umlein~\protect\cite{Bluemlein} (dashed line) 
at $\mu^2=10$ GeV$^2$
and of Kwiecinski, Martin, Stasto~\protect\cite{Martin_Stasto} (dotted line).
  }}\label{gluon}
\end{figure}

\begin{figure}[htb]
  \vspace*{2mm}
\epsfig{figure=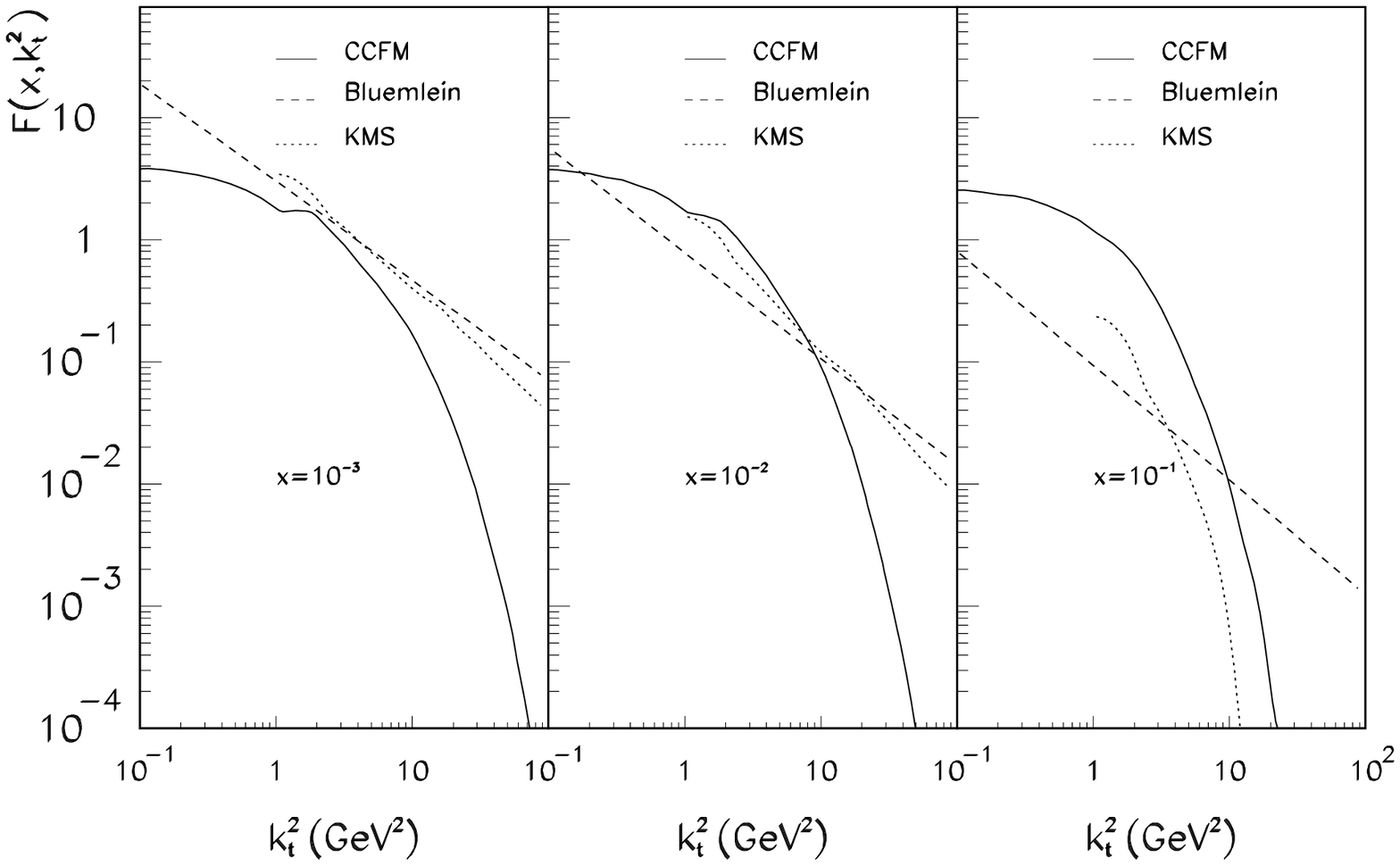,
width=17cm,height=10cm}
\caption{{\it
The gluon density ${\cal A}(x,k_t^2,\bar{q}^2)$ (solid line) obtained from
a solution of the CCFM equation~\protect\cite{CASCADE} as a function of
$k_t^2$ for different values of $x$ (at $\bar{q}^2=10$ GeV$^2$).
Also shown for comparison is the unintegrated gluon density function
${\cal F}(x,k^2_{t},\mu^2)$
 according
to the parameterization of Bl\"umlein~\protect\cite{Bluemlein} (dashed line) 
at $\mu^2=10$ GeV$^2$
and of Kwiecinski, Martin, Stasto~\protect\cite{Martin_Stasto} (dotted line).
  }}\label{gluon_kt}
\end{figure}

The parton evolution at small values of $x$ is believed to be best described by
the CCFM evolution equation \cite{\CCFM}, 
which for $x \to 0$ is equivalent to
the BFKL evolution equation \cite{\BFKL} and for large $x$ reproduces the
standard DGLAP equations. The CCFM evolution equation takes coherence effects
of the radiated gluons into account via angular ordering.
In \cite{CASCADE} it is shown that a very good description of the inclusive
structure function $F_2(x,Q^2)$ and the production of forward jets in DIS,
which are believed to be a prominent signature of small $x$ parton dynamics,
can be obtained from the CCFM evolution equation.
 The main important point there
was the treatment of the non-Sudakov form factor, which suppresses radiation at
small values of $x$. 
In Fig.~\ref{gluon} we show the gluon density obtained from this solution of the
CCFM equation~\cite{CASCADE}  
as a function of $x$ for 
different values of $k_t^2$ at $\bar{q}^2 = 10$ GeV$^2$.
In Fig.~\ref{gluon_kt} we show the gluon density
as a function of $k_t^2$ for 
different values of $x$ at $\bar{q}^2 = 10$ GeV$^2$. 
\par
For comparison, we also use the results of a BFKL-like
 parameterization of
the unintegrated gluon distribution 
${\cal F}(x,k^2_{t},\mu^2)$, according to the prescription in~\cite{Bluemlein}. 
The proposed method lies upon a straightforward perturbative solution of the
BFKL equation where the collinear gluon density $x\,G(x,\mu^2)$
from the standard GRV set~\cite{GRVc} is used as
the boundary condition in the integral form (\ref{kt}).
Technically, the unintegrated gluon density is calculated as a convolution
of collinear gluon density with universal weight factors~\cite{Bluemlein}:
\begin{equation} \label{conv}
 {\cal F}(x,k_t^2,\mu^2) = \int_x^1
 {\cal G}(\eta,k_t^2,\mu^2)\,
 \frac{x}{\eta}\,G(\frac{x}{\eta},\mu^2)\,d\eta,
\end{equation}
\begin{equation} \label{J0}
 {\cal G}(\eta,k_t^2,\mu^2)=\frac{\bar{\alpha}_s}{xk_t^2}\,
 J_0(2\sqrt{\bar{\alpha}_s\ln(1/\eta)\ln(\mu^2/k_t^2)}),
 \qquad k_t^2<\mu^2,
\end{equation}
\begin{equation}\label{I0}
 {\cal G}(\eta,k_t^2,\mu^2)=\frac{\bar{\alpha}_s}{xk_t^2}\,
 I_0(2\sqrt{\bar{\alpha}_s\ln(1/\eta)\ln(k_t^2/\mu^2)}),
 \qquad k_t^2>\mu^2,
\end{equation}
where $J_0$ and $I_0$ stand for Bessel functions (of real and imaginary
arguments, respectively), and $\bar{\alpha}_s=3{\alpha}_s/\pi$.
The latter parameter is connected with the Pomeron trajectory intercept:
$\Delta=\bar{\alpha}_s4\ln{2}$ in the LO and 
$\Delta=\bar{\alpha}_s4\ln{2}-N\bar{\alpha}_s^2$ in the NLO approximations,
respectively, where $N$ is a number \cite{Salam1}. In the following we use
$\Delta=0.35$.
\par
The presence of the two different  parameters, $\mu^2$ and $k^2_{t}$, in
eq.(\ref{conv})  
for  unintegrated gluon distribution 
${\cal F}(x,k^2_{t},\mu^2)$ refers  to the fact that
the evolution of parton densities is done in two steps. 
First  the DGLAP scheme is applied to evolve the
structure function from $Q^2_0$ to $\mu^2$ within the 
collinear approximation.
After that  eqs.(\ref{conv})-(\ref{I0}) are used to develop the 
parton transverse momenta $k_t^2$.
This is in contrast to the CCFM evolution, where the 
evolution of ``longitudinal"
and ``transverse" components occurs simultaneously.
\par
From Figs.~\ref{gluon} and~\ref{gluon_kt} we see that
the BFKL approach gives much harder $k_t$ spectrum than the CCFM approach. 
However, it has been argued extensively 
in the literature~\cite{Martin_Stasto,Martin_Sutton2,Martin}, that in BFKL a
so-called ``consistency constraint" should be applied, so simulate at least a
part of the large next-to-leading corrections. For comparison we also show in 
Figs.~\ref{gluon} and~\ref{gluon_kt} the unintegrated gluon distribution from
Kwiecinski, Martin, Stasto~\cite{Martin_Stasto}\footnote{A. Stasto kindly 
provided  us (H.J.) with the program.}. This shows that the shape of the
distributions from BFKL including the ``consistency constraint" is similar to
the one obtained from CCFM and that the gluon density is stronger suppressed at
large $k_t$ as compared to the approach in \cite{Bluemlein}.
 Unfortunately we could not use the the gluon
distribution from~\cite{Martin_Stasto}, because it started only at $k_t^2>1$.
\section{Predictions for $D^*$ meson production at HERA}
We have used the hadron level Monte Carlo program
\CASCADE~ described in \cite{CASCADE} to predict
the cross section for $D^*$ photo-production at HERA energies. The unintegrated
gluon distribution was obtained from the solution of the CCFM equation described
in \cite{CASCADE}. The scale in $\alpha_s$ was set to $k_t^2$ in the parton
evolution and we used $\Lambda_{QCD}^{(4)} = 0.2$ GeV.
 For the hard scattering
 the off-shell matrix element for heavy quarks (including the heavy quark
mass with $m_c=1.5$ GeV) 
as described in \cite{off_shell_me} are used together with
 the one-loop expression for
$\alpha_s$ with $k_t^2$ (of the gluon
entering the hard scattering) as the scale.
 The complete initial state cascade 
is simulated via a backward evolution as described in \cite{CASCADE}. The
hadronization was performed by the Lund string fragmentation 
JETSET~\cite{\JETSET}. The 
Peterson  function with $\epsilon = 0.06$ was used for the charm quark
fragmentation. 
\begin{figure}[htb]
  \vspace*{2mm}
\epsfig{figure=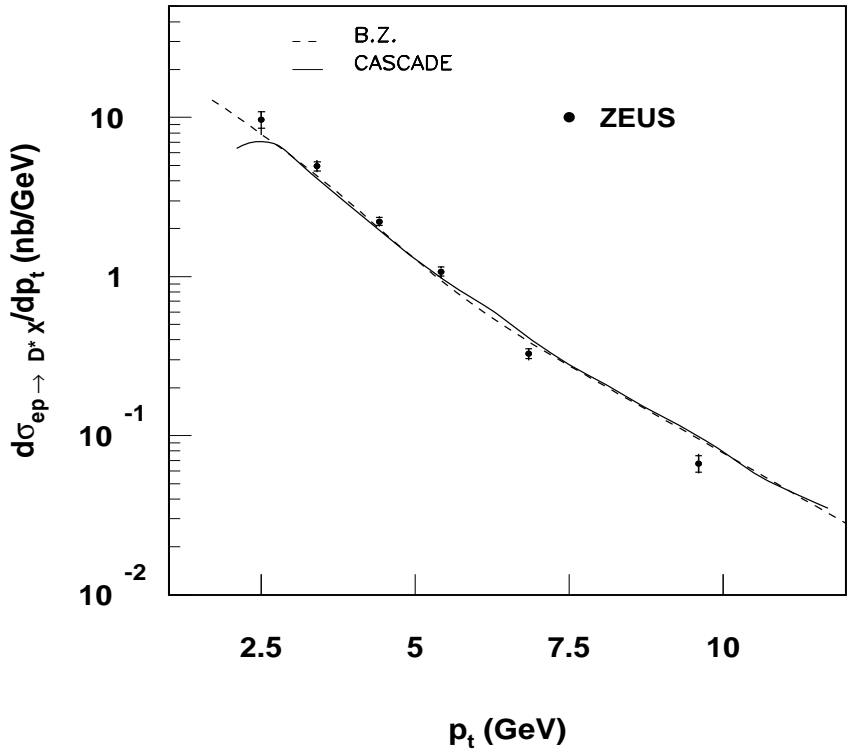,
width=15cm,height=12cm}
\caption{{\it The differential cross section $d \sigma /dp_t^{D^*}$ for 
$Q^2<1$ GeV$^2$.
The solid line shows the prediction from the \CASCADE~ Monte Carlo and the dashed
line is the calculation of \protect\cite{baranov_zotov1}.
 The data point are from \protect\cite{ZEUS_dstar}.  }}\label{dstar_pt}
\end{figure}
\par
In Fig.~\ref{dstar_pt} we show the prediction of $D^*$ production as a 
function of the transverse momentum  $p_t^{D^*}$ using the \CASCADE~
Monte Carlo described above and compare it with the measurement of the ZEUS
collaboration~\cite{ZEUS_dstar}. We observe a rather good description of the
$p_t$ spectrum.
\par
\begin{figure}[htb]
  \vspace*{2mm}
\epsfig{figure=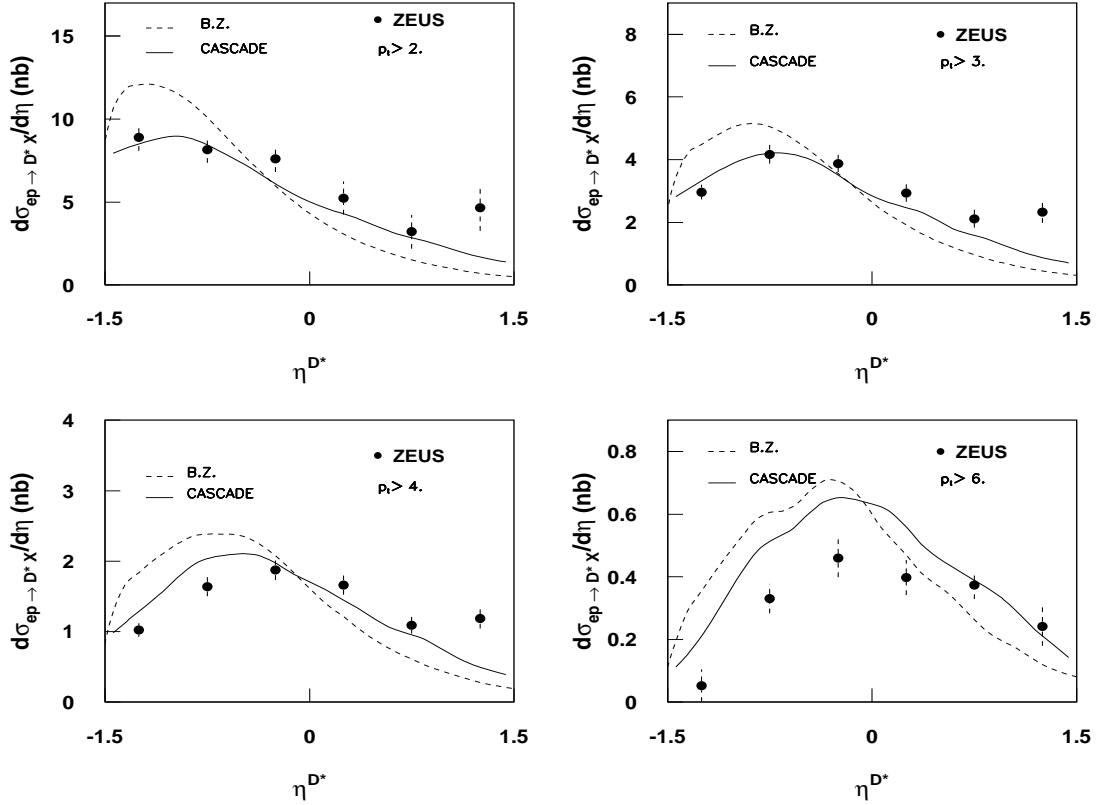,
width=17cm,height=12cm}
\caption{{\it The differential cross section $d \sigma /d\eta^{D^*}$ for 
$Q^2<1$ GeV$^2$ for different regions of $p_t^{D^*}$.
The solid line show the prediction from the \CASCADE~ Monte Carlo and the dashed
line is the calculation of \protect\cite{baranov_zotov1}.
 The data point are from \protect\cite{ZEUS_dstar}.
  }}\label{dstar_eta}
\end{figure}
In Fig.~\ref{dstar_eta} we show the $D^*$ cross section as a function of the
pseudo-rapidity $\eta^{D^*}$
for different regions in $p_t$. Also here we observe a  
good description of the experimental data points.  Here the main important point
is the cross section at values of $\eta^{D^*} > 0.5$. 
\par
For comparison we used the results of a calculation of~\cite{baranov_zotov1}
 with a parameterization  for
the unintegrated gluon distribution 
${\cal F}(x,k^2_{t},\mu^2)$, according to the prescription 
of~\cite{Bluemlein}. Here the scale 
$m_t^2$
is used in $\alpha_s$
The results are also shown in Figs.~\ref{dstar_pt} and 
\ref{dstar_eta}. In general both prediction agree rather nicely. The differences
observed are entirely due to the different behavior of the unintegrated gluon
distribution as a function of $x$ and $k_t^2$. In addition in the CCFM approach
angular ordering and the maximum angle allowed for any emission plays a
important role. The results presented here are similar to the ones obtained from
a full NLO calculation. In the SHA approach the gluon entering the hard
interaction is off-mass shell, which is a similar situation as in a 
NLO calculation where the propagators in the 3 parton final states are fully
considered. This is in contrast to the LO DGLAP (collinear) case, where the
gluon is always treated on-mass shell. However if, in LO DGLAP,
 heavy flavor excitation is
included (via resolved photon processes) then again a similar situation occurs.
It was found in \cite{ZEUS_dstar} that including heavy flavor excitation in LO
Monte Carlo programs, lead to better description of the data. 
The semi-hard approach presented here shows, that a good description of also the
photo-production of heavy flavor can be achieved in a theoretically consistent
way, without including artificially large intrinsic transverse momenta, or heavy
flavor excitation.
\section{Conclusions}
We have shown that using 
a unintegrated gluon distribution obtained from a solution of
the CCFM evolution equation, that describes the structure function $F_2(x,Q^2)$
and the production of forward jets in DIS at HERA, we can also describe the
cross sections of inclusive $D^{*\pm}$ meson production measured at HERA.
Within the
semi-hard approach the measured cross section as a function of  $p_{t}$
and $\eta_{D^*}$ can be nicely described. 
The results are similar to NLO calculations.
The shape of the gluon $k_t$ distribution is driven by the BFKL or CCFM
evolution equations. This shows that there is no place to include any 
artificially
large intrinsic transverse momentum distribution of parton inside the proton.
\par
It is also interesting to note, that within the semi-hard approach, heavy flavor
excitation in the photon is consistently included, by the fact that a gluon
radiated close to the quark box can have a transverse momentum larger that 
of the quarks. 
\section{Acknowledgments}
We are grateful to M. Ryskin and Y. Shabelski for many discussions about the
semi-hard approach.
We thank  L. Gladilin  for many discussions about the ZEUS $D^*$ data.

\end{document}